\documentclass[runningheads]{llncs}
\usepackage[utf8]{inputenc}
\usepackage[small]{caption}
\usepackage{graphicx}
\usepackage{amsmath}
\usepackage{tabularx}
\usepackage{booktabs}

\usepackage{subcaption}
\usepackage{array}
\newcolumntype{H}{>{\setbox0=\hbox\bgroup}c<{\egroup}@{}}
\usepackage{multirow}

\usepackage{enumitem}

\usepackage{times}

\usepackage{geometry}
\usepackage[numbers,sort&compress]{natbib}
\usepackage[hyphens,spaces]{url}
\usepackage{hyperref,xcolor}  
\usepackage{accsupp}
\newcommand{\shortcite}[1]{\citep{#1}}

\newtheorem{hyp}{Hypothesis}
\usepackage{microtype}

\usepackage{xcolor}

\title{Towards Faster Reasoners By Using Transparent Huge Pages}

\author{Johannes K. Fichte\inst{1} \and Norbert Manthey \and Julian Stecklina \and Andr\'e Schidler\inst{2}}

\institute{TU Dresden \and TU Wien}

\definecolor{dblue}{rgb}{0,0,0.54}

\DeclareMathOperator{\var}{var}

\begin{document}

\maketitle

\begin{abstract}
  Various state-of-the-art automated reasoning (AR) tools are widely
  used as backend tools in research of knowledge representation and
  reasoning as well as in industrial applications. In testing and
  verification, those tools often run continuously or nightly.
  In this work, we present an approach to reduce the runtime of AR
  tools by 10\% on average and up to 20\% for long running tasks.  Our
  improvement addresses the high memory usage that comes with the data
  structures used in AR tools, which are based on conflict driven
  no-good learning.
  We establish a general way to enable faster memory access by using
  the memory cache line of modern hardware more effectively.
  Therefore, we extend the standard C library (glibc) by dynamically
  allowing to use a memory management feature called huge pages.
  Huge pages allow to reduce the overhead that is required to
  translate memory addresses between the virtual memory of the
  operating system and the physical memory of the hardware. 
  In that way, we can reduce runtime, costs, and energy consumption of
  AR tools and applications with similar memory access patterns simply
  by linking the tool against this new glibc library when compiling
  it. In every day industrial applications this easily allows to be
  more eco-friendly in computation.
  To back up the claimed speed-up, we present experimental results for
  tools that are commonly used in the AR community, including the
  domains ASP, BMC, MaxSAT, SAT, and SMT.

\end{abstract}

\section{Introduction}

Very recently, Vardi~\cite{Vardi20} directed attention to the fact that traveling to numerous conferences unsurprisingly results in a bad carbon-dioxide footprint.
While in this case impact to the environment is
immediate and obvious, there are many hidden factors that impact the environment. We often neglect the factors relating to the computation for combinatorial problems, as they seem to yield only a small improvement.
This applies in particular to state-of-the-art solvers in
combinatorial problem solving and \emph{automated reasoning (AR)},
such as ASP, \#SAT, MaxSAT, MUS, SAT, and SMT solvers.
The underlying algorithms for those reasoners are often based on a
technique called conflict driven no-good learning (CDNL) for which it
is well-known that their efficiency highly depends on their memory
consumption~\cite{BiereHeuleMaarenWalsh09}.  A common explanation is
that advanced data structures benefit from improved memory
access. Examples are data structures for learnt clauses, two watched
literals, and linear lookup tables.  The impact is particularly large
when the implementations respect hardware memory caches.

On modern systems, accessible memory is virtual and handled by a
physical memory management unit (MMU). The mapping between virtual and
physical memory is stored in page tables by the MMU.  In order to
reduce access time, recently used mappings are stored in a
\emph{translation lookaside buffer (TLB)}.
Since combinatorial reasoners often consume a significant amount of
memory, avoiding cache misses and page translation
failures can considerably improve the performance of a SAT
solver~\cite{HolldoblerMantheySaptawijaya10}.
Because CDNL-based solvers often form the core of other reasoning
tools, improving the memory behavior of modern solvers can
considerably speed up their execution. Even if the improvement
involves only small factors the practical impact can be
huge. Consider an industrial test case that runs each day 5 hours
during the night. If we are able to reduce that runtime just by 10\%,
it results in 30 minutes less computation time per night. Overall this
sums up to 10 hours of saved energy consumption per month. In
practice, this efficiency improvement saves money and allows more
eco-friendly computation.

\subsection*{New Contribution}
In this paper, we introduce a simple and transparent approach to
effectively reduce the number of TLB misses in order to speed up the
execution of modern memory dependent solvers, in particular, unit
propagation or sometimes also called Boolean constraint propagation.
We employ a Linux memory management feature called \emph{transparent
  huge pages (THP)}, which reduces the overhead of virtual memory
translations by using larger virtual memory page sizes~\cite{kernel_thp}.
Our approach is based on modifying the standard C library (glibc),
which is the default standard library in Linux
systems~\cite{ArnoldEtAl19}. Whenever a solver allocates memory, we
make sure that we additionally give the operating system kernel advice
about the use of the memory (madvise).
This feature can then be used for a solver simply by recompiling it
and statically linking it against our modified glibc.
In that way, we obtain a significant speed-up on benchmarks in model
checking of up to 15\% and for most other solvers up to~10\%.
The approach is based on a hardware feature and thus generalizable to
other operating systems and CPU architectures supporting large page
sizes.

\medskip
\noindent Our advances summarize as follows:
\begin{enumerate}
\item We propose an easily accessible way to reduce the number of TLB
  misses in combinatorial memory-dependent solvers by patching the
  glibc such that our modifications can be activated or deactivated
  at runtime.

\item We provide a build system to easily patch glibc and statically
  link a solver against the patched version. Our system is based on
  a setup that uses OS-level virtualization
  (docker)~\cite{Solomon19} and is available to all modern Linux
  systems. We already provide various pre-compiled state-of-the-art
  reasoning tools.
\item We carry-out extensive benchmarks and present detailed results
  for various reasoning tools. %
\end{enumerate}

\subsection*{Related Work}

Chu, Hardwood, and Stuckey~\cite{ChuHarwoodStuckey09a} as well as
Holldobler, Manthey, Saptawijaya~\cite{HolldoblerMantheySaptawijaya10}
considered cache utilization in SAT solvers and illustrated how a
resource-unaware SAT solver can be improved by utilizing the cache
sensibly, resulting in reasonable speed-ups.
The effect of huge pages has already been widely investigated in the field of
operating systems, e.g.,~\cite{PanwarPrasadGopinath18}, mostly with a strong
focus on database systems.
However, to the best of our knowledge the is no overall study on
combinatorial problem solving and automated reasoning.
Recent research considered benchmarking system
tools~\cite{KouweAndriesseBos18a}, selecting benchmarks to tune
solvers~\cite{HoosKaufmannSchaub13a}, and treating input benchmarks
for benchmarking~\cite{BiereHeule19a}.
These topics are orthogonal to our work. In contrast, we consider computational
resources and memory management of solvers, in particular, its
effect on the runtime.
Bornebusch, Wille, and Drechsler~\cite{BornebuschWilleDrechsler17a}
analyzed the memory footprint of SAT solvers and tried to improve
them. However, they did not consider propagation and its data
structures, which is reasonable from a complexity point of view due to
large formula sizes.

\section{Modern CDNL-based Solvers and Memory}
Before we present our advances, we give a brief explanation on how
modern SAT solvers are implemented and introduce components and
mechanisms that are relevant for memory access.  As many reasoners are
based on SAT
technology,~e.g.,~\cite{BiereHeuleMaarenWalsh09,clasp,Voronkov14a},
core concepts are very similar for various reasoners.
First, we define (propositional) formulas and their evaluation in the
usual way and assume familiarity with standard notations, including
satisfiability. For basic literature, we refer to introductory
work~\cite{KleineBuningLettman99}. We consider a universe~$U$ of
propositional variables. A \emph{literal} is a variable or its
negation and a \emph{clause} is a finite set of literals. A (CNF)
\emph{formula} is a finite set of clauses. A \emph{(partial) truth
  assignment} is a mapping $\tau: \var(X) \rightarrow \{0,1\}$ defined
for a set~$X \subseteq U$ of variables.
For $x \in X$, we put $\tau(\neg x) = 1 - \tau(x)$.
For a formula~$F$, we abbreviate by $\var(F)$ the variables that
occur in~$F$.
We say that a truth assignment~$\tau$ \emph{satisfies} a clause~$C$,
if for at least one literal~$\ell \in C$ we have $\tau(\ell) = 1$.  We
say that a truth assignment~$\tau$ falsifies a clause~$C$, if it
assigns all its literals to $0$.
We call a clause~$C$ \emph{unit} if~$\tau$ assigns all but one literal
to~$0$.
A truth assignment~$\tau$ is \emph{satisfying} if for each
clause~$C \in F$, the truth assignment~$\tau$ satisfies~$C$.

\subsection{Why SAT solvers are fast?}
So far, there are two main contributing factors to advances in the efficiency of modern SAT solvers: (i) theoretical improvements in terms of more advanced algorithms and heuristics and (ii) algorithm engineering in terms of data structures.
The core algorithm that drives search in modern SAT solvers is based
on \emph{conflict driven no-good learning (CDNL)} or also known as
\emph{conflict driven clause learning
  (CDCL)}~\cite{MarquessilvaSakallah99,GomesSelmanCrato97}, which was
widely extended by search
heuristics~\cite{BiereHeuleMaarenWalsh09,KatebiSakallahMarquessilva11}
and simplification techniques during
search~\cite{JarvisaloHeuleBiere12}.
A key technique is \emph{unit propagation}, which aims at finding
clauses where all literals but one are already assigned and then
setting the remaining literal to a value that satisfies the
clause. Unit propagation is responsible for the vast majority of the overall
runtime even in modern
solvers~\cite{KatebiSakallahMarquessilva11}. Hence, algorithm
engineering and efficient data structures are essential for practical
solving,~i.e., the \emph{two-watched-literal} scheme for
unit propagation~\cite{MoskewiczEtAl01} and fast lookup tables, which are also
important for the used heuristics and learning techniques.
The watched literal scheme reduces the number of steps in the
algorithm and memory accesses, but decreases the efficiency of the
memory access~\cite{HolldoblerMantheySaptawijaya10}. Still, this
results in a considerable overall runtime
improvement~\cite{KatebiSakallahMarquessilva11}.
While lookup tables provide fast accesses of relevant clauses, they
result in a much higher memory footprint and may yield 
unpredictable memory access~\cite{HolldoblerMantheySaptawijaya10}.

\subsection{CPUs, Virtual Memory, and Paging}

Modern \emph{operating systems (OSes)} provide the concept of
\emph{virtual memory} to applications. Thereby, the OS releases
software developers from worrying about the actual physical memory
layout and also allows for overcommitting resources.  Virtual memory
is managed at the granularity of \emph{pages}. A page is a contiguous
block of memory in the virtual address space. The OS can map a page to
a \emph{page frame}, which is a corresponding location in physical
memory.
On the Intel Architecture~\cite{intelsdm}, page tables %
describe the mapping from pages to page frames and thus virtual to
physical addresses. On 64-bit Intel systems, these page tables are trees
with a depth of commonly up to four levels for $2^{48}$-byte address
spaces.

Walking these data structures to provide a translation for each memory
access is infeasible, because it would add one page table read per level
for each intended memory access. Instead, processors take advantage of
spatial and temporal locality of memory accesses and cache
translations in Translation Lookaside Buffers (TLBs).
An Intel Skylake system has two-levels of TLBs and the unified L2 TLB
can hold 1536 entries~\cite{skylake-architecture}.
Other recent CPUs have similar specifications.
With 4~KiB pages, this translates to holding translations for 6~MiB of virtual memory in the TLB.
A straight-forward way to increase the capacity of the TLB is for the
processor architecture to allow for larger page sizes. On Intel 64-bit
systems, in addition to 4~KiB pages, the system also supports 2~MiB
and 1~GiB pages. While 1~GiB pages have (few) dedicated TLB entries,
2~MiB pages share the same entries as 4~KiB entries in the TLB on
Skylake.

\begin{figure}[t]%
\centering
\resizebox{1\textwidth}{!}{%
  \includegraphics{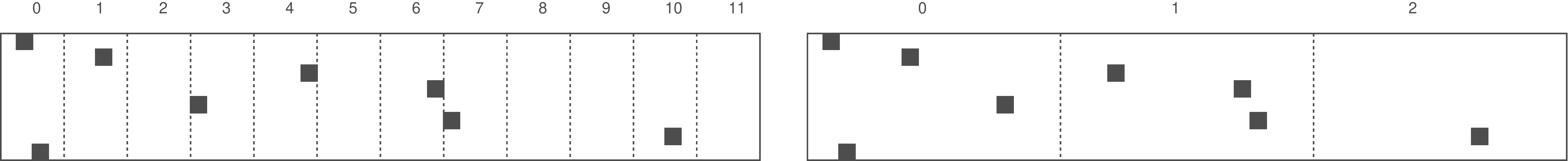}
}
\caption{This figure illustrates how pages cover a sequence of memory accesses for two different sizes of pages for a given amount of memory.
}
\label{fig:thp}
\end{figure}

To benefit from large pages, the OS needs to make them accessible  to applications~\cite{Arcangeli10a}.  The main problem is
constantly defragmenting memory to have contiguous free memory from
which large pages can be allocated~\cite{navarro02}.
Figure~\ref{fig:thp} illustrates the usage of memory with pages of
different sizes. When using larger pages, less pages are required to cover the same area of memory, and hence, less TLB entries are occupied.
In more detail, the black boxes in Figure~\ref{fig:thp} illustrate a
sequence of accesses, which start at the top and flow to the bottom.
While for larger pages (right) it is sufficient to memorize the
translation for three pages, smaller pages require seven pages
(left). In case the TLB can only hold four entries, the entry of
Page~0 would be evicted before it can be re-used to access the same
clause again.  When using larger pages, less initial translations have
to be done, and only three pages are required to perform all accesses.

\subsection{Large Pages in Linux}
Large page sizes are supported in Linux by a feature called
\emph{transparent huge pages (THP)}, which offers both implicit and explicit
use of large page sizes and was introduced in Linux 2.6.38~\cite{kernel_thp}.
If THP is enabled, memory does not have to be statically provisioned
for applications to use large pages, which is a clear advantage over
previous attempts involving large pages~\cite{ParkKimYeom19a}.
Instead, the system is continuously compacting memory to free up
contiguous space to allocate large pages. The Linux kernel can then,
depending on the system configuration, transparently allocate large
pages for applications. If intended, a system administrator can still
additionally provision large pages manually.
THP can be globally enabled or be configured as an opt-in
feature. Both mechanisms degrade gracefully when no large pages are
available and will instead back memory using the standard page size.
When THP is configured for opt-in, an application can use the system call ``madvise''
with the ``MADV\_HUGEPAGE'' flag to mark memory regions as eligible. If
this is done for virtual memory regions that have not been backed by
physical memory yet,~e.g., directly after a ``mmap'' call, the kernel
will try to allocate a large page on first access to this memory.
Otherwise, the kernel will occasionally scan virtual memory that is
eligible for THP to create large pages.
One downside of THP is that the kernel has to run scan and compact
operations. Linux allows to configure this behavior to mitigate the
impact by paying the cost for scanning and compacting at allocation
time instead of doing it as a background job.

\subsection{The Effect of THP}
System workloads are known to speed up with huge pages.  However, it
may also result in reproducibility decreases, as huge pages have to be
enabled in the kernel and globally for all applications on the
system~\cite{ParkKimYeom19a}.  Hence, it is often recommended to use
small pages for benchmarking.  Unfortunately, the StarExec
cluster~\cite{StumpSutcliffeTinelli14} has enabled THP by default for
all executed programs, which comes with the mentioned downsides.
In case of virtualized machines, using small pages can result in
almost 50\% of the runtime being spent in address
translation~\cite{KwonYuPeter16a}.  Using 2M pages in both the guest
and host reduces this value to about 4\%.
We expect similar savings for tools that are run in a virtualized
environment, as virtualization typically uses huge pages internally.
In the following of the paper, we investigate the actual effect on a
bare metal setup for our experimental work.

\section{TLB Misses in SAT Solvers}
Typically, SAT solvers do not exhibit the memory access locality that caches or
TLB are optimized for.
While previous works considered
caches~\cite{ChuHarwoodStuckey09a,HolldoblerMantheySaptawijaya10},
memory translation and the TLB have not been taken into account.
Hence, we focus on memory accesses in the most time consuming part of
SAT solvers: Boolean constraint propagation or also called unit
propagation. 

Assume that a formula~$F$ and a partial truth assignment~$\tau$ is given
and a two watched literal data structure is
used~\cite{MoskewiczEtAl01}.  Briefly, \emph{unit propagation} works as outlined in
Listing~\ref{alg:bcp}.
Initially, for each clause~$C \in F$, one selects two 
literals from~$C$, which are not non-falsified by truth assignment~$\tau$. Then, the
truth assignment~$\tau$ is extended by setting additional
variables. Since assigning a literal~$\ell \in C$ such that
$\tau(\ell)=1$ results in clause~$C$ that is satisfied, which in turn
allows to remove clause~$C$ from the considered clauses right away,
the only interesting case is if truth assignment~$\tau$ sets
literal~$\ell \in C$ such that $\tau(\ell)=0$. In that case, the
clause~$C$ might be falsified and be involved in a conflict or have
unassigned literals, which can be used to imply the truth value of
other literals.
Then, UnitPropagate checks every clause~$C$ that contains a literal which might be falsified during propagation.
Therefore, the list~$P$ of literals to propagate  is traversed (Line~B1), and each watch list~$L_p$ for literal~$p$ is processed (Line~B3).
Then, each clause~$C$ in list~$L_p$ contains $\neg p$, so that the new state of clause~$C$ has to be evaluated by processing the other literals in clause~$C$.
Hence there are two cases: either (i)~clause $C$ is satisfied by another literal, or (ii) clause~$C$ contains another literal~$x$ that is not yet falsified by the truth assignment~$\tau$ (Line~B5).
Then, we watch literal~$x \in C$ for being set to false instead of $\neg p$, and consequently have to update list~$L_p$ (Line~B6) and list~$L_{\neg x}$ (Line~B7).
Otherwise, clause~$C$ might be a unit clause (Line~B8) or might be falsified by truth assignment~$\tau$ (Line~B9).
In both cases, clause~$C$ can remain in list~$L_p$.

\begin{table}[t]
\centering
\def\tablename{Listing}

   \sf\small
  \def\L#1{\raise .2ex\hbox{\tiny\tt #1}&}
  \def\C#1{\hspace{1.5em}\hfill\bf//\small#1}%
  \def\S#1{\mbox{\emph{#1}}}
  \def\K#1{\textbf{#1}}
  \def\N{\\[.4ex]}                                          
  \def\I{\hspace{1em}}
  \begin{tabularx}{0.725\textwidth}{r@{\hspace{5ex}}l@{\hspace{3ex}}l}
  \toprule
  \L{}$\S{UnitPropagate}$ (formula $F$, truth assignment~$\tau$, literals $P$, watch lists $L$) \N
  \midrule
  \L{B1}  \I \K{while} the list~$P$ of literals to propagate is not empty \C{compute closure} \N
  \L{B2}  \I\I pick $p \in P$, and remove from $P$ \C{typically DFS} \N
  \L{B3}  \I\I access watch list $L_p$ of clauses $\neg{p} \in C$ \C{propagate} \N
  \L{B4}  \I\I \K{for all} clauses $C$ in $L_p$: \N
  \L{B5}  \I\I\I \K{if} $C \neq \emptyset$, $x \in C$, and $x$ not falsified in assignment~$\tau$ \C{watchable literal} \N
  \L{B6}  \I\I\I\I remove $C$ from $L_p$ \C{maintain lists} \N
  \L{B7}  \I\I\I\I add $C$ to watch list $L_{\neg x}$ for $\neg x$ \C{maintain lists} \N
  \L{B8}  \I\I\I \K{else if} $C = (x)$ unit, extend $P$ and $\tau$ with $x$ \C{unit rule} \N
  \L{B9}  \I\I\I \K{else if} $C$ is falsified, trigger \K{conflict analysis($\tau$, $C$)} \C{conflict} \N
  \bottomrule
  \end{tabularx}
\vspace{0.5em}
\caption{Pseudo code for an implementation of unit propagation with the watched literal scheme. The state of the solver holds the formula~$F$ as watch lists $L$, one list $L_x$ for each literal $x$, as well as a truth assignment~$\tau$, and the list $P$ of literals  to propagate. The result of the algorithm with either be an extension of the truth assignment, or a tuple truth assignment~$\tau$ and a conflict clause that is falsified by truth assignment~$\tau$.}
\label{alg:bcp}
\end{table}

When considering the memory access pattern, the unit propagation algorithm has the following properties:
the literals in list~$P$ are not easily determined in advance.
Hence, accesses in Line B3 to load the list are hard to predict.
One could reduce the memory accesses in Lines~B3 and B4 by pre-fetching data from memory in advance. This has been proposed in previous works~\cite{HolldoblerMantheySaptawijaya10}.
Accessing the clauses in Line~B4 are hard to predict, as the order of the clauses in list~$L_p$ changes. In more detail, in Line~B6 some clauses are removed and in Line~B8 or B9 others are kept.
To improve the access behavior in Line B5, Een and Sorensson~\cite{EenSorensson04a} proposed for MiniSAT~2.1 an optimization to
avoid the access of clause~$C$ for the satisfied case. There,
another literal of clause~$C$ is stored in list~$L_p$. Since its introduction, blocking literals are commonly used in
most modern solvers.  Accessing literal~$x$ in Line B5 is also unpredictable,
as literal order in clauses also changes. Typically, the two watched
literals are the first two stored literals and they change whenever the
clause is moved to another list.

We suspect that unit propagation is the major source of memory accesses, as most
run time is spend in unit propagation, and many different memory locations,~i.e.,
clauses are accessed non-linearly during unit propagation. This drives us to the
following hypothesis:

\begin{hyp}\label{hyp:bcpTLBoffender} %
  Accessing clauses during unit propagation as well as updating and accessing
  watch lists has the highest impact on TLB misses.
\end{hyp}

We support our hypothesis by the following observation.
The two watched literals data structure allows to keep the number of overall accesses
low, but trades it for higher memory footprint with additional data
structures and lists. 
Further, the memory accesses for (i)~clause to access next,
(ii)~literals of a clause to watch next, or (iii)~list to place it are
difficult to predict. Hence, Lines B3, B4, and B7 are prime candidates
to access memory locations that have not been accessed recently, and
hence, are not cached, nor served with current TLB entries.

\subsection{Analyzing Unit Propagation}
To back up Hypothesis~\ref{hyp:bcpTLBoffender} with data, we analyze the distribution of TLB misses in the SAT solvers MiniSat and Glucose\footnote{We use a sampling approach of CPU performance counters for TLB misses with the system tool perf.}.
When running MiniSat~\cite{EenSorensson04a}, we observe that 90\% of
TLB misses occur in unit propagation; thereof, about 10\% when moving clause to
another (unpredictable) watch list and about 80\% when accessing the
first literal of the next watched clause.  This data matches the
assumption that Line B3 and B4 are responsible for most of the TLB misses.
In addition, moving clauses to new watch lists contributes another
10\%.
When running Glucose version~4.2.1~\cite{glucose421}, we can see similar results. 90\% of the TLB misses happen in unit propagation.
In the modern solver Glucose, unit propagation is split into (i) propagating binary clauses that contributed 5\,\% of all TLB misses, (ii) propagating during learned clause minimization~\cite{ijcai2017-98} that contributes about 20\,\%, and (iii) propagating larger clauses and pushing them to watch lists that consume the majority of the TLB misses.
Empirical observations for these two solvers confirm our hypothesis, unit propagation is the
major source for TLB misses.  The random memory accesses to check the
next clause in the list for being unit, which can have an arbitrary
memory location, as well as putting clauses into another watch list,
are the major contributors.  Unit propagation is responsible for a
large fraction of the run time of SAT solvers, which is actually spend
in address translation.  Biere~\shortcite{Biere16a} places in his
solver watch lists and its clauses closer to each other, to avoid TLB
misses related to Line B4.
Here, we present an orthogonal approach to avoid TLB misses, which
allows to improve the implementation~\cite{Biere16a} further and can
be applied to many other solvers.

\subsection{Counter the Unit Propagation Implementation}
We believe that additional data structure improvements along~\cite{ChuHarwoodStuckey09a} are hardly feasible.
Clauses would have to be even more compact and the changes require a huge effort for a single solver~\cite{Biere2014LingelingEA}.
Changes to the underlying algorithms likely result in reduced
performance and require tuning parameters again.  On that account, we
propose a general approach to THP in the following, which can also be
easily used by other tools.

\section{Improving Unit Propagation with THP via Madvise}

Modern Linux distributions provide native support for transparent huge
pages. Usually, the systems allow the superuser to define the behavior
via the configuration file
``\url{/sys/kernel/mm/transparent_hugepage/enabled}'' whose values
``always'' or ``never'' apply to all running processes.
Because there might be applications running on the host that would
suffer from larger pages, THP is usually disabled on physical
systems and it is not advised to set the value to always.
Fortunately as described above, Linux also allows to set the value using the madvise system call. While this sounds fairly trivial, it requires (i) lots of
manual adaptions of the source code to mark memory regions as eligible
and in turn makes the implementations of solvers (ii) fairly
incomparable on an algorithmic level.
In the following, we suggest an easily accessible way to reduce the number of TLB misses
in combinatorial memory-dependent solvers.

\subsection{Using More Huge Pages}
In the previous section, we explained that using more huge pages seems
a reasonable approach to speed up the memory access of a modern
solvers. This can be obtained by running a madvise system call to
instruct the kernel to use transparent huge pages of 2M whenever the
solver allocates memory.
Then, we align all requested memory to 2M addresses and increase the
size of the reservation accordingly, so that huge pages can actually
be used.  If we would not do so, two memory requests of the application
can be in the middle of an 2M page, which results in not using a huge
page. Compared to the system setting, this change results in using one more
huge page per misaligned memory request.

\subsection{Patching the Standard C Library (glibc)}
In order to provide a transparent way to various solver developers,
and offering a way for algorithms engineering to consider the effect
of transparent huge pages on many AR tools, we want to avoid manual
source code adaptions as much as possible.
To this end, we put our focus on the standard C system library glibc,
which already provides standard functions to access the system
memory. The library is used in Linux to compile most of the solvers.
Instead of modifying the source code of various solvers, we implement
the above mentioned ideas into glibc\footnote{Our latest
  implementation is publicly available at
  \url{https://github.com/conp-solutions/thp}.  Note that we will upstream
  the patches of glibc along this publication to merge it into glibc
  if accepted by the glibc maintainers.
}.
Whenever a certain runtime flag is activated, our modified glibc takes
care of the above mentioned changes.
The current approach uses a system environment variable
(\texttt{\nolinkurl{GLIBC_THP_ALWAYS=1}}) that can be specified before calling a
program.
This way we allow setting the flag for a specific program instead of
all running programs. Globally enabling transparent huge pages for all
running applications on the system is usually forbidden both in
industry and academia by administrators due to a variety of potential
side effects, which might slow down a variety of programs.  If the
flag is not set, we disable the use of huge pages. The additional
cost, compared to glibc, is a single if-statement.  We implemented our
patches into glibc~2.23~\cite{ArnoldEtAl19} and 
enable the feature without any source code modification of the various
solvers themselves.
In that way it is entirely sufficient for the user of a solver to
recompile the solver and link it against our modified glibc.

\subsection{Huge Pages in a Solver}
In order to use the feature, there are two ways to proceed: (i) \emph{link}
the solver \emph{statically} against our modified glibc or (ii) patch the
system glibc and then \emph{dynamically link} the solver against the new
glibc.
We provide an easy and accessible way for the former, since patching
the system glibc is usually considered problematic due to side effects
and as it requires superuser permissions, which makes it very unlikely
that actual users of the reasoning tools will use this feature.
We introduce a virtual environment that allows for easy compiling of
the solver, in order to avoid problematic setups of a new secondary
glibc.

Our system is based on the OS-level virtualization
docker~\cite{Solomon19}, which isolates running programs entirely from
each other. 
Docker itself is available on all modern OSes and allows to deliver
software in packages, which are called containers. A running container
is entirely isolated from one another and can bundle its own
software. We use this to not interfere with the system glibc. But we
do not publish only a docker container, instead we provide the scripts
to build containers in which the compilation then runs.
The user just needs to install docker and we provide the tooling to
link a solver with THP support. Along, we give many exemplary
scripts to highlight how to run the tools and various pre-compiled
state-of-the-art solvers.

\section{Experimental Evaluation}
We conducted a series of experiments using standard benchmark sets for
various reasoning tools. All benchmark sets and our results are
publicly available.
To represent many fields, we selected various benchmarks and tools.

\subsection{Benchmarked Solvers and Instances}
In our experimental work, we present results for recent versions of
publicly available SAT solvers: Glucose~4.2.1~\cite{glucose421},
lingeling~\cite{Biere17a}, MapleLCMDistChronoBTDL
(winner2019)~\cite{MapleLCMDistChronoBTDL},
MergeSat~\cite{mergesat}, MiniSAT~\cite{EenSorensson04a},
plingeling~\cite{Biere17a}.
For SAT, we selected the recommended benchmark for tool tuning, and
compare MiniSat and MergeSat from the above set of SAT tools again. %
For answer set programming (ASP), we used clasp~\cite{clasp} and a
benchmark that has been shown to be recommended for benchmark
selection~\cite{HoosKaufmannSchaub13a}.
From software model checking (SWMC), we use CBMC~\cite{ckl2004}, which
uses a single call to a SAT solver.  As SWMC benchmark, we use the
benchmark provided when introducing the LLBMC
tool~\cite{10.1007/978-3-642-27705-4_12}.
As another group, we collected tools, which use incremental SAT
solvers as a backend.  For hardware model checking (HWMC), we use the
bounded model checker aigbmc~\cite{Biere-FMV-TR-11-2},
with an unrolling limit of 100, and use the benchmark of the deep
bound track of the HWMC Competition of 2017~\cite{BiereVanDijkHeljanko-FMCAD17}.
For optimization, we use the MaxSAT solver
Open-WBO~\cite{10.1007/978-3-319-09284-3_33}, which uses the SAT solver Glucose as
a backend.  As MaxSAT benchmark, we picked the weighted partial maxsat
formulas from 2014, to make sure the incremental interface is actually
used.  Finally, we consider muser-2~\cite{DBLP:journals/jsat/BelovM12}, which
computes a minimal unsatisfiable subformula (MUS) from a CNF formula,
and use the group MUS benchmark from the MUS competition
2011\footnote{MUS benchmarks are available at
\href{http://www.cril.univ-artois.fr/SAT11/bench/SAT11-Competition-GMUS-SelectedBenchmarks.tar}{\nolinkurl{cril.univ-artois.fr/SAT11}}}.

\begin{figure}
  \centering
    \resizebox{0.8\columnwidth}{!}{\includegraphics{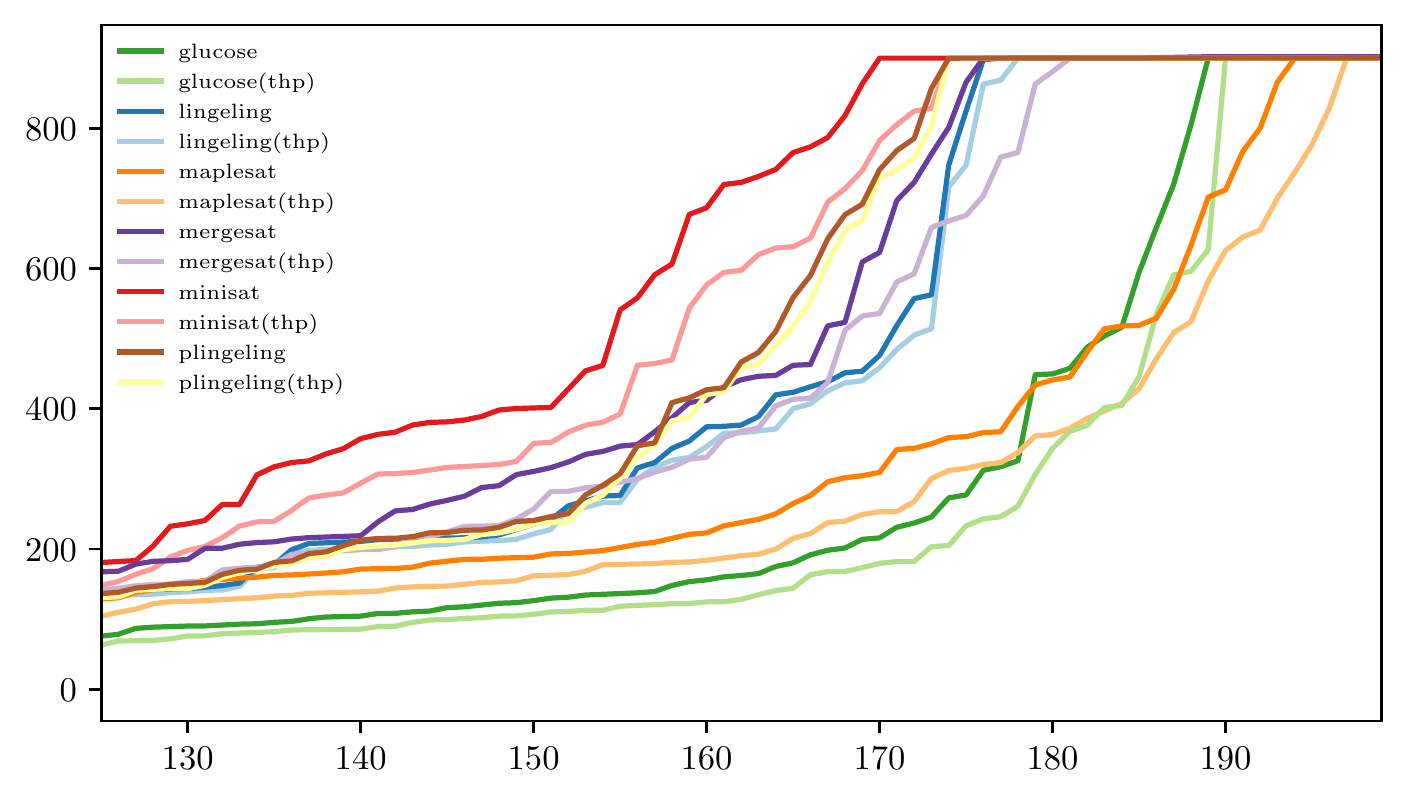}}
    \caption{Runtime for the SAT solvers on all considered
      instances. The x-axis refers to the number of instances and the
      y-axis depicts the runtime sorted in ascending order for each
      solver individually.}
    \label{fig:cactus}
\end{figure}

\begin{table}
\centering
\begin{tabular}{Hl@{\hskip 4ex}||@{\hskip 2ex}r@{\hskip 2ex}|@{\hskip 2ex}r@{\hskip 3ex}r@{\hskip 4ex}Hr@{\hskip 2ex}|@{\hskip 2ex}l@{\hskip 3ex}l@{\hskip 4ex}Hr}
\toprule
{} &      solver & $\#$ &  $t_n$ & $t_{thp}$ &  $f$ & $s$ &       $\text{TLB}_n$ &   $\text{TLB}_{thp}$ & $s_{tlb}$ & $r_{tlb}$\\
\midrule
1 &     glucose &   189 &  4.58 &     3.72 &  1.23 &  18.70 & 2.60E+11 &  6.71E+09 &  38.67 &   2.59 \\
2 &   lingeling &   177 &  6.18 &     5.91 &  1.05 & 4.76 & 4.93E+10 &  5.13E+08 &  96.28 &   1.04 \\
0 &    winner19 &   194 &  7.46 &     6.24 &  1.20 & 16.67 & 3.29E+11 &  1.35E+10 &  24.40 &   4.10 \\
3 &    mergesat &   176 &  7.31 &     6.22 &  1.17 & 14.53 & 3.02E+11 &  1.36E+10 &  22.21 &   4.50 \\
4 &     minisat &   170 &  7.26 &     6.09 &  1.19 &  15.97 & 2.75E+11 &  2.97E+09 &  92.39 &   1.08 \\
\bottomrule
\end{tabular}
\vspace{1em}
\caption{Overview on the speed-up between solving when using THP for SAT solvers. $\#$ counts the number of solved instances.  $t$ contains the runtime in hours, $s$ the saved runtime in \%,~i.e., $s = 100 - (t_n/t_{thp}\cdot 100)$ factor, and $\text{TLB}$ the TLB load misses. We distinguish by non-THP and THP by $\cdot_n$ and $\cdot_{thp}$, respectively. $r_{tlb}$ summarizes the \% of TLB misses over the original TLB misses,~i.e., $r_{tlb} = \text{TLB}_{thp} / \text{TLB}_{n} \cdot 100$.}
\label{fig:sat:thp}
\end{table}

\subsection{Boolean Satisfiability (SAT)}
For SAT solvers we carried out an extensive study on a cluster.

\subsubsection{Measure, Setup, and Resource Enforcements.}
Our results were gathered on a cluster of RHEL~7.4 Linux machines with
kernel 3.10.0-693 on GCC~4.8.5-16. %
We evaluated the solvers on machines with
two Intel Xeon E5-2680v3 CPUs of 12
physical cores each at 2.50GHz base frequency. We forced the
performance governors to 2.5Ghz~\cite{HackenbergEtAl19a} and disabled
multi-threading.
The machines are equipped with 64GB main memory of which 60.5GB are freely available to programs.
We compare wall clock time and number of timeouts.
However, we avoid IO access on the CPU solvers whenever
possible,~i.e., we load instances into the RAM before we start
solving.
We run at most 5 solvers on one node, set a timeout of 900 seconds,
and limited available RAM to~8 GB per instance and solver.
We follow standard guidelines for
benchmarking~\cite{KouweAndriesseBos18a}.

\subsubsection{SAT Results.}
Figure~\ref{fig:cactus} illustrates the runtime results of the solvers
in different configurations in a cactus-like plot.
Table~\ref{fig:sat:thp} gives an overview on the number of solved
instances for each solver with and without THP. Note that we report in
this table only on instances that have been solved by non-THP and THP.
The results show that a solver with activated THP solves overall more
instances that without THP.
When considering runtime, the configurations that employ THP solve the
considered instances faster. Runtime improvements range from a small
improvement for lingeling, which was about 0.27h faster, up to more
than one hour faster for mergesat, minisat, and winner2019. In terms
of factor of saved runtime hours, we can see that the solvers that
employ THP are up to almost $19\%$ faster.
The number of TLB misses that we observed reduce up to $2$ orders,
namely, $r_{tlb} = \text{TLB}_{thp} / \text{TLB}_{n}$ goes down to
$1\%$ for lingeling and similar for glucose and minisat.
When we consider the number of solved instances glucose solves 5
instances less than winner2019. However, glucose solves the
instances in~4.58 hours while winner2019 solves them in 7.46 hours.

\subsubsection{Discussion and Summary.}
The observed results above confirm our hypotheses. Throughout our
experiments, the number of TLB misses goes significantly down for all
considered solvers.  For lingeling and minisat it even reduces to 1\%
of the original number of misses. Since unit propagation is a major source of
memory accesses and major cause of a high number of TLB misses and
responsible of a large part of the solving time, the reduced TLB
misses also yield a speed up in the overall runtime. Our approach
makes all solvers faster and even allows glucose to be almost as fast
as winner19 without spending a significant amount of time in
optimizing the solver itself.
Only very few benchmarks can be solved in less than 1MB or 10MB
memory. Hence, using larger pages results in less TLB misses and in
faster execution

\subsection{Other Reasoners.}

We believe that the THP approach does not only boost SAT solvers, but other reasoners as well.
On that account, we run additional experiments on tools that are either based on SAT solvers or implemented closely to the CDNL algorithm.
To broaden the applicability, we also consider tools that use SAT solvers via their incremental interface~\cite{EenSorensson04a}.
As the SAT calls in these tools are shorter, we expect that the benefit of using THP is smaller.

\begin{hyp}\label{hyp:incrementalBenefitsLess}
When using incremental SAT solvers inside a reasoner, the benefit of THP is smaller.
\end{hyp}

\paragraph{Measure, Setup, and Resource Enforcements.}
To support Hypothesis~\ref{hyp:incrementalBenefitsLess}, we run a
second analysis. To make sure the above results are not CPU and OS
dependent, we used a second environment of same architecture and
repeat the results for MiniSat and MergeSat.
The computer has an Intel Core i5-2520M CPU running at
2.50GHz, with an Ubuntu 16.04 and Linux~4.15, using 5GB as memory
limit and the 900 seconds as timeout per instance.

\begin{table}[t]
  \centering
    \begin{tabular}{@{\hskip 2em}l@{\hskip 1em}l@{\hskip 2em}|@{\hskip 2em}r@{\hskip 1em}r@{\hskip 1em}r@{\hskip 2em}}
      Category & Tool     & $t_n$  & $t_{thp}$      & $s[\%]$ \\
      \toprule
      SAT      & MiniSat  & 8.17 & 7.03  & 13.99 \\
      SAT      & MergeSat & 7.94 & 6.90  & 13.13 \\
      \midrule
      ASP      & clasp    & 3.66 & 3.29  & 10.18 \\
      MaxSAT   & open-wbo & 1.19 & 1.09  & 8.49 \\
      MUS      & muser2   & 4.18 & 3.97  & 5.16 \\
      \midrule
      HWMC     & aigbmc   & 0.89 & 0.86  & 4.11 \\
      SWMC     & cbmc     & 0.23 & 0.22  &  2.76 \\
      \bottomrule
    \end{tabular}
    \vspace{1em}
    \caption{Overview on the runtime of various reasoners with(out) THP evaluated on their respective competition benchmarks. $t_n$ and $t_{thp}$ represent the runtime   of the reasoner in hours and $s$ represents the saved runtime in \%.}
    \label{table:summaryTHPRuntimeOtherTools}
\end{table}

\subsubsection{Results.} Table~\ref{table:summaryTHPRuntimeOtherTools}
states the results for the considered tools and benchmarks. 
To measure the speed-up, we show only instances that could also be
solved by the variant without THP.  When using THP, we can solve the
same instances within the timeout, usually a few more.
First, we can see a similar improvement as in the previous setting where we used the same architecture, but different hardware.
The
improvement when using THP for tools with a single calls is similarly
high as presented above,~i.e., SAT as well as ASP show
improvements above 10\%.  Only cbmc from SWMC is an outlier, which
might be related to it memory usage.
For tools that use incremental SAT solver as a backend, the improvement range from 4\% from HWMC to 8\% in MaxSAT.
The low speedup can be explained with their memory usage: over each
benchmark, cbmc's memory usage is rather low,~i.e., the median memory
footprint is 8.8MB.  For all the other tools and categories, the
memory footprint is higher,~e.g., the median for ASP is 28.8MB and for
SAT 123.3MB.  The tools with incremental SAT backends also consume
more memory than cbmc: MaxSAT 21.3MB, HWMC 164MB, and MUS 298.1MB.  As
expected, tools with a higher memory footprint result in a higher
speed-up due to transparent huge pages.

\section{Conclusion and Future Work}
Although reasoners solve NP-hard problems, they are used across the
research community to solve many tasks in artificial
intelligence. Reasoners are also employed in industry to verify properties,
generate tests, or run similar tasks.
In this paper, we introduced a simple and transparent approach to
effectively speed up memory access of reasoners by reducing the number
of TLB misses, which in turn allows to significantly improve their
runtime.
Our approach is based on a modification of glibc, which is the C
standard library of the GNU Project and widely used in Linux for C and
C++ programs.
A user of a reasoner can benefit from our improvement simply by
recompiling his favorite reasoner and enabling the feature by an
environment flag when invoking it.
Our experiments confirmed that an application save 25\% runtime on
certain instances and on average more than 10\%.
Since the tools are often also used for long running jobs, for example
in systems biology or verification, we can save a significant amount of
runtime and hence save energy and money.
In that way, the number of solved instances might not always be the
right measure to evaluate a reasoner.

We believe that our approach can also be very beneficial other tools
in automated reasoning, simply because there are many memory incentive applications
that have simply not been tuned to reduce the number of TLB misses by
using THP.
One such domain might be graph algorithms, which can have random
memory access patterns if the underlying data structure is updated
often.
In the future, we are interested in the influence of THP to various
other domains and to increase the amount of tested benchmarks and
reasoning tools.
We hope that this opens up both theoretical and practical research on
more general algorithm engineering techniques solvers.
\bibliographystyle{named}
\bibliography{thp_arxiv2020}

\begin{thebibliography}{10}
\providecommand{\url}[1]{\texttt{#1}}
\providecommand{\urlprefix}{URL }
\providecommand{\MyDOI}[2]{\BeginAccSupp{E=Digital Object
  Identifier}\EndAccSupp{}\href{#1}{\texttt{#2}}}

\bibitem{sc2019proceedings}
Proceedings of {SAT} {Race} 2019 : {Solver} and {Benchmark} {Descriptions},
  Department of Computer Science Report Series, vol. B-2019-1. University of
  Helsinki, Helsinki, Finland (2019)

\bibitem{Arcangeli10a}
Arcangeli, A.: Transparent hugepage support. In: KVM forum. vol.~9 (2010)

\bibitem{ArnoldEtAl19}
Arnold, R.S., Brown, M., Eggert, P., Jelinek, J., Kuvyrkov, M., Myers, J.,
  O'Donell, C., Oliva, A., Schwab, A.: The gnu c library (glibc).
  \url{https://www.gnu.org/software/libc/} (2019)

\bibitem{glucose421}
Audemard, G., Simon, L.: {Glucose in the SAT Race 2019}.
  \cite{sc2019proceedings}, pp. 19--20

\bibitem{DBLP:journals/jsat/BelovM12}
Belov, A., Marques{-}Silva, J.: Muser2: An efficient {MUS} extractor. J. on
  Satisfiability, Boolean Modeling and Computation  \textbf{8}(3/4),  123--128
  (2012)

\bibitem{Biere2014LingelingEA}
Biere, A.: Lingeling essentials, a tutorial on design and implementation
  aspects of the the sat solver lingeling. In: POS@SAT (2014)

\bibitem{Biere16a}
Biere, A.: {Splatz, Lingeling, Plingeling, Treengeling, YalSAT Entering the SAT
  Competition 2016}. In: Balyo, T., Heule, M., J{\"a}rvisalo, M. (eds.)
  Proc.~of {SAT Competition} 2016 -- Solver and Benchmark Descriptions.
  Department of Computer Science Series of Publications B, vol. B-2016-1, pp.
  44--45. University of Helsinki (2016)

\bibitem{Biere17a}
Biere, A.: {CaDiCaL, Lingeling, Plingeling, Treengeling, YalSAT Entering the
  SAT Competition 2017}. In: Balyo, T., Heule, M., J{\"a}rvisalo, M. (eds.)
  Proc.~of {SAT Competition} 2017 -- Solver and Benchmark Descriptions.
  Department of Computer Science Series of Publications B, vol. B-2017-1, pp.
  14--15. University of Helsinki (2017)

\bibitem{BiereVanDijkHeljanko-FMCAD17}
Biere, A., van Dijk, T., Heljanko, K.: Hardware model checking competition
  2017. In: Stewart, D., Weissenbacher, G. (eds.) Formal Methods in
  Computer-Aided Design, {FMCAD} 2017, Vienna, Austria, October 02-06, 2017.
  p.~9. {IEEE} (2017)

\bibitem{Biere-FMV-TR-11-2}
Biere, A., Heljanko, K., Wieringa, S.: {AIGER 1.9} and beyond. Tech. Rep.~11/2,
  Institute for Formal Models and Verification, Johannes Kepler University,
  Altenbergerstr. 69, 4040 Linz, Austria (2011)

\bibitem{BiereHeule19a}
Biere, A., Heule, M.: The effect of scrambling {CNFs}. In: Berre, D.L.,
  J\"arvisalo, M. (eds.) Proceedings of Pragmatics of SAT 2015 and 2018. EPiC
  Series in Computing, vol.~59, pp. 111--126. EasyChair (2019)

\bibitem{BiereHeuleMaarenWalsh09}
Biere, A., Heule, M., van Maaren, H., Walsh, T. (eds.): Handbook of
  Satisfiability, Frontiers in Artificial Intelligence and Applications,
  vol.~185. IOS Press, Amsterdam, Netherlands (Feb 2009)

\bibitem{BornebuschWilleDrechsler17a}
Bornebusch, F., Wille, R., Drechsler, R.: Towards lightweight satisfiability
  solvers for self-verification. In: Proceedings of the 7th International
  Symposium on Embedded Computing and System Design (ISED'17). pp.~1--5 (Dec
  2017),
  \MyDOI{https://doi.org/10.1109/ISED.2017.8303924}{10.1109/\allowbreak{}ISED.2017.8303924}

\bibitem{ChuHarwoodStuckey09a}
Chu, G., Harwood, A., Stuckey, P.: Cache conscious data structures for
  {Boolean} satisfiability solvers. J. on Satisfiability, Boolean Modeling and
  Computation  \textbf{6},  99--120 (02 2009),
  \MyDOI{https://doi.org/10.3233/SAT190064}{10.3233/\allowbreak{}SAT190064}

\bibitem{ckl2004}
Clarke, E., Kroening, D., Lerda, F.: A tool for checking {ANSI-C} programs. In:
  Jensen, K., Podelski, A. (eds.) Tools and Algorithms for the Construction and
  Analysis of Systems (TACAS 2004). Lecture Notes in Computer Science,
  vol.~2988, pp. 168--176. Springer (2004)

\bibitem{EenSorensson04a}
E{\'e}n, N., S{\"o}rensson, N.: An extensible {SAT}-solver. In: Giunchiglia,
  E., Tacchella, A. (eds.) Proceedings of the 6th International Conference on
  Theory and Applications of Satisfiability Testing (SAT'03). pp. 502--518.
  Springer Verlag (2003)

\bibitem{GomesSelmanCrato97}
Gomes, C., Selman, B., Crato, N.: Heavy-tailed distributions in combinatorial
  search. In: Smolka, G. (ed.) Proceedings of the 3rd International Conference
  on Principles and Practice of Constraint Programming (CP'97). Lecture Notes
  in Computer Science, vol.~1330, pp. 121--135. Springer Verlag, Linz, Austria
  (1997),
  \MyDOI{https://doi.org/10.1007/BFb0017434}{10.1007/\allowbreak{}BFb0017434}

\bibitem{HackenbergEtAl19a}
Hackenberg, D., Sch{\"o}ne, R., Ilsche, T., Molka, D., Schuchart, J., Geyer,
  R.: An energy efficiency feature survey of the intel haswell processor. In:
  Lalande, J.F., Moh, T. (eds.) Proceedings of the 17th International
  Conference on High Performance Computing \& Simulation (HPCS'19) (2019)

\bibitem{HolldoblerMantheySaptawijaya10}
H{\"o}lldobler, S., Manthey, N., Saptawijaya, A.: Improving resource-unaware
  {SAT} solvers. In: Ferm{\"u}ller, C.G., Voronkov, A. (eds.) Proceedings of
  the 16th International Conference on Logic for Programming, Artificial
  Intelligence, and Reasoning (LPAR'16). Lecture Notes in Computer Science,
  vol.~6397, pp. 519--534. Springer Verlag, Dakar, Senegal (2010),
  \MyDOI{https://doi.org/10.1007/978-3-642-16242-8_26}{10.1007/\allowbreak{}978-\allowbreak{}3-\allowbreak{}642-\allowbreak{}16242-\allowbreak{}8\_\allowbreak{}26}

\bibitem{HoosKaufmannSchaub13a}
Hoos, H.H., Kaufmann, B., Schaub, T., Schneider, M.: Robust benchmark set
  selection for boolean constraint solvers. In: Proceedings of the 7th
  International Conference on Learning and Intelligent Optimization (LION'13).
  Lecture Notes in Computer Science, vol.~7997, pp. 138--152. Springer Verlag,
  Catania, Italy (Jan 2013), revised Selected Papers

\bibitem{Solomon19}
Hykes, S., et~al.: Docker ce. \url{https://github.com/docker/docker-ce} (2019)

\bibitem{intelsdm}
Intel: Intel{\textregistered} 64 and {IA}-32 {A}rchitectures {S}oftware
  {D}eveloper's {M}anual (2019), order Number: 325462-069US

\bibitem{JarvisaloHeuleBiere12}
J{\"a}rvisalo, M., Heule, M., Biere, A.: Inprocessing rules. In: Gramlich, B.,
  Miller, D., Sattler, U. (eds.) Automated Reasoning, Lecture Notes in Computer
  Science, vol.~7364, pp. 355--370. Springer Verlag (2012),
  \MyDOI{https://doi.org/10.1007/978-3-642-31365-3_28}{10.1007/\allowbreak{}978-\allowbreak{}3-\allowbreak{}642-\allowbreak{}31365-\allowbreak{}3\_\allowbreak{}28}

\bibitem{KatebiSakallahMarquessilva11}
Katebi, H., Sakallah, K.A., Marques-Silva, J.P.: Empirical study of the anatomy
  of modern {SAT} solvers. In: Sakallah, K.A., Simon, L. (eds.) Proceedings of
  the 14th International Conference on Theory and Applications of
  Satisfiability Testing (SAT'11), Lecture Notes in Computer Science,
  vol.~6695, pp. 343--356. Springer Verlag, Ann Arbor, MI, USA (June 2011),
  \MyDOI{https://doi.org/10.1007/978-3-642-21581-0_27}{10.1007/\allowbreak{}978-\allowbreak{}3-\allowbreak{}642-\allowbreak{}21581-\allowbreak{}0\_\allowbreak{}27}

\bibitem{clasp}
Kaufmann, B., Gebser, M., Kaminski, R., Schaub, T.: clasp -- a conflict-driven
  nogood learning answer set solver. \url{http://www.cs.uni-potsdam.de/clasp/}
  (2015)

\bibitem{KleineBuningLettman99}
Kleine~B{\"u}ning, H., Lettman, T.: Propositional logic: deduction and
  algorithms. Cambridge University Press, Cambridge, New York, NY, USA (1999)

\bibitem{MapleLCMDistChronoBTDL}
Kochemazov, S., Zaikin, O., Kondratiev, V., Semenov, A.:
  {MapleLCMDistChronoBT-DL, duplicate learnts heuristic-aided solvers at the
  SAT Race 2019}.  \cite{sc2019proceedings}, pp. 24--24

\bibitem{KouweAndriesseBos18a}
van~der Kouwe, E., Andriesse, D., Bos, H., Giuffrida, C., Heiser, G.:
  Benchmarking crimes: An emerging threat in systems security. CoRR
  \textbf{abs/1801.02381} (2018), \url{http://arxiv.org/abs/1801.02381}

\bibitem{KwonYuPeter16a}
Kwon, Y., Yu, H., Peter, S., Rossbach, C.J., Witchel, E.: Coordinated and
  efficient huge page management with ingens. In: Proceedings of the 12th
  USENIX Conference on Operating Systems Design and Implementation (OSDI'16).
  pp. 705---721. USENIX Association, Savannah, GA, USA (2016),
  \MyDOI{https://doi.org/10.5555/3026877.3026931}{10.5555/\allowbreak{}3026877.3026931}

\bibitem{ijcai2017-98}
Luo, M., Li, C.M., Xiao, F., Many{\`a}, F., L{\"u}, Z.: An effective learnt
  clause minimization approach for {CDCL} {SAT} solvers. In: Proceedings of the
  Twenty-Sixth International Joint Conference on Artificial Intelligence,
  {IJCAI-17}. pp. 703--711 (2017),
  \MyDOI{https://doi.org/10.24963/ijcai.2017/98}{10.24963/\allowbreak{}ijcai.2017/\allowbreak{}98}

\bibitem{mergesat}
Manthey, N.: {MergeSat}.  \cite{sc2019proceedings}, pp. 29--30

\bibitem{MarquessilvaSakallah99}
Marques-Silva, J., Sakallah, K.: {GRASP}: a search algorithm for propositional
  satisfiability. IEEE Transactions on Computers  \textbf{48}(5),  506--521
  (May 1999),
  \MyDOI{https://doi.org/10.1109/12.769433}{10.1109/\allowbreak{}12.769433}

\bibitem{10.1007/978-3-319-09284-3_33}
Martins, R., Manquinho, V., Lynce, I.: Open-wbo: A modular maxsat solver,. In:
  Sinz, C., Egly, U. (eds.) Theory and Applications of Satisfiability Testing
  -- SAT 2014. pp. 438--445. Springer International Publishing, Cham (2014)

\bibitem{10.1007/978-3-642-27705-4_12}
Merz, F., Falke, S., Sinz, C.: Llbmc: Bounded model checking of c and c++
  programs using a compiler ir. In: Joshi, R., M{\"u}ller, P., Podelski, A.
  (eds.) Verified Software: Theories, Tools, Experiments. pp. 146--161.
  Springer Berlin Heidelberg, Berlin, Heidelberg (2012)

\bibitem{MoskewiczEtAl01}
Moskewicz, M.W., Madigan, C.F., Zhao, Y., Zhang, L., Malik, S.: {Chaff}:
  Engineering an efficient {SAT} solver. In: Rabaey, J. (ed.) Proceedings of
  the 38th Annual Design Automation Conference (DAC'01). pp. 530--535. Assoc.
  Comput. Mach., New York, Las Vegas, Nevada, USA (2001),
  \MyDOI{https://doi.org/10.1145/378239.379017}{10.1145/\allowbreak{}378239.379017}

\bibitem{navarro02}
Navarro, J., Iyer, S., Druschel, P., Cox, A.: Practical, transparent operating
  system support for superpages. SIGOPS Oper. Syst. Rev.  \textbf{36}(SI),
  89--104 (Dec 2003),
  \MyDOI{https://doi.org/10.1145/844128.844138}{10.1145/\allowbreak{}844128.844138},
  this paper describes Super Page implementation in FreeBSD. It also has
  performance nubmers, but really ancient ones. They roughly match the SAT
  solver performance improvements, though.

\bibitem{PanwarPrasadGopinath18}
Panwar, A., Prasad, A., Gopinath, K.: Making huge pages actually useful. In:
  Bianchini, R., Sarkar, V. (eds.) Proceedings of the 23rd ACM International
  Conference on Architectural Support for Programming Languages and Operating
  Systems (ASPLOS'18). pp. 679--692. Assoc. Comput. Mach., New York,
  Williamsburg, VA, USA (Mar 2018),
  \MyDOI{https://doi.org/10.1145/3173162.3173203}{10.1145/\allowbreak{}3173162.3173203}

\bibitem{ParkKimYeom19a}
Park, S., Kim, M., Yeom, H.Y.: {GCMA}: Guaranteed contiguous memory allocator.
  IEEE Transactions on Computers  \textbf{68}(3),  390--401 (Mar 2019),
  \MyDOI{https://doi.org/10.1109/TC.2018.2869169}{10.1109/\allowbreak{}TC.2018.2869169}

\bibitem{StumpSutcliffeTinelli14}
Stump, A., Sutcliffe, G., Tinelli, C.: Starexec: A cross-community
  infrastructure for logic solving. In: Demri, S., Kapur, D., Weidenbach, C.
  (eds.) Proceedings of the 7th International Joint Conference on Automated
  Reasoning (IJCAR'14). Lecture Notes in Computer Science, vol.~8562, pp.
  367--373. Springer Verlag, Vienna, Austria (Jul 2014),
  \MyDOI{https://doi.org/10.1007/978-3-319-08587-6_28}{10.1007/\allowbreak{}978-\allowbreak{}3-\allowbreak{}319-\allowbreak{}08587-\allowbreak{}6\_\allowbreak{}28},
  held as Part of the Vienna Summer of Logic, VSL 2014.

\bibitem{kernel_thp}
Torvalds, L.: kernel.org: Transparent hugepage support.
  \url{https://www.kernel.org/doc/Documentation/vm/transhuge.txt} (May 2017)

\bibitem{Vardi20}
Vardi, M.Y.: Publish and perish. Communications of the ACM  \textbf{63}(1), ~7
  (2020), \MyDOI{https://doi.org/10.1145/3373386}{10.1145/\allowbreak{}3373386}

\bibitem{Voronkov14a}
Voronkov, A.: Avatar: The architecture for first-order theorem provers. In:
  Biere, A., Bloem, R. (eds.) Proceedings of the 26th International Conference
  on Computer Aided Verification {CAV'14}. Lecture Notes in Computer Science,
  vol.~8559, pp. 696--710. Springer Verlag (2014), held as Part of the Vienna
  Summer of Logic (VSL).

\bibitem{skylake-architecture}
Wikichip, C.: {Skylake (client) -- Microarchitectures -- Intel}.
  \url{https://en.wikichip.org/wiki/intel/microarchitectures/skylake_(client)}
  (2020)

\end{thebibliography}
\end{document}